\documentclass[12pt]{article}
\usepackage{times}
\usepackage{geometry}
\usepackage[utf8]{inputenc}
\usepackage{enumitem,amssymb}
\usepackage{ragged2e}

\usepackage[english]{babel} % Specify a different language here - english by default
\usepackage{natbib}
\usepackage{multicol}
\usepackage{pdfpages}
\usepackage{enumitem}
\usepackage[leftcaption]{sidecap}
\usepackage{wrapfig}

\newcommand{\apjl}{ApJL}

\newcommand{\ao}{Appl. Optics}
\newcommand{\aap}{A\&A}

\newlist{thematic}{itemize}{8}
\setlist[thematic]{label=$\square$}
\usepackage{pifont}

\begin{document}
\raggedright
\huge
Astro2020 Science White Paper \linebreak

New Frontiers for Terrestrial-sized to Neptune-sized Exoplanets In the Era of Extremely Large Telescopes \linebreak
\normalsize

\noindent \textbf{Thematic Areas:} \hspace*{60pt} $\boxtimes$ Planetary Systems \hspace*{10pt} $\square$ Star and Planet Formation \hspace*{20pt}\linebreak
$\square$ Formation and Evolution of Compact Objects \hspace*{31pt} $\square$ Cosmology and Fundamental Physics \linebreak
  $\square$  Stars and Stellar Evolution \hspace*{1pt} $\square$ Resolved Stellar Populations and their Environments \hspace*{40pt} \linebreak
  $\square$    Galaxy Evolution   \hspace*{45pt} $\square$             Multi-Messenger Astronomy and Astrophysics \hspace*{65pt} \linebreak
  
\textbf{Principal Author:}

Name: Ji Wang \& Michael R. Meyer
 \linebreak						
Institution: The Ohio State University \& University of Michigan 
 \linebreak
Email: wang.12220@osu.edu \& mrmeyer@umich.edu
 \linebreak
Phone: 614-292-1773 \& 734-764-7846 
 \linebreak
 
\textbf{Co-authors:} (names and institutions)
  \linebreak
Alan Boss, Carnegie Institution\\
Laird Close, University of Arizona\\
Thayne Currie, NASA-Ames Research Center\\
Diana Dragomir, MIT\\
Jonathan Fortney, UC Santa Cruz\\
Eric Gaidos, University of Hawaii at Manoa\\
Yasuhiro Hasegawa, JPL/Caltech\\
Irina Kitiashvili, NASA Ames\\
Quinn Konopacky, UC San Diego\\
Chien-Hsiu Lee, National Optical Astronomy Observatory\\
Nikole K. Lewis, Cornell University\\
Michael Liu, University of Hawaii\\
Roxana Lupu, BAER Institute\\
Dimitri Mawet, Caltech\\
Carl Melis, UC San Diego\\
Mercedes L${\rm \acute{o}}$pez-Morales, Smithsonian Astrophysical Observatory\\
Caroline V. Morley, University of Texas at Austin\\
Chris Packham, University of Texas at San Antonio\\
Eliad Peretz, NASA, Goddard Space Flight Center\\
Andy Skemer, UC Santa Cruz\\
Mel Ulmer, Northwestern University\\
\pagebreak

\textbf{Abstract:} Surveys reveal that terrestrial- to Neptune-sized planets (1 $< R <$ 4 R$_{\rm{Earth}}$) are the most common type of planets in our galaxy. Detecting and characterizing such small planets around nearby stars holds the key to understanding the diversity of exoplanets and will ultimately address the ubiquitousness of life in the universe.  The following fundamental questions will drive research in the next decade and beyond:  (1) how common are terrestrial to Neptune-sized planets within a few AU of their host star, as a function of stellar mass?  (2) How does planet composition depend on planet mass, orbital radius, and host star properties?  (3) What are the energy budgets, atmospheric dynamics, and climates of the nearest worlds? Addressing these questions requires:  a) diffraction-limited spatial resolution; b) stability and achievable contrast delivered by adaptive optics; and c) the light-gathering power of extremely large telescopes (ELTs), as well as multi-wavelength observations and all-sky coverage enabled by a comprehensive US ELT Program. Here we provide an overview of the challenge, and promise of success, in detecting and comprehensively characterizing small worlds around the very nearest stars to the Sun with ELTs. This white paper extends and complements the material presented in the findings and recommendations published in the National Academy reports on Exoplanet Science Strategy and Astrobiology Strategy for the Search for Life in the Universe.

\pagebreak

\section{Scientific Justification}\label{sec:scijust}
\begin{wrapfigure}[39]{r}{0.55\textwidth}
    \vspace{-0.5in}
    \centering
    \includegraphics[width=0.55\textwidth]{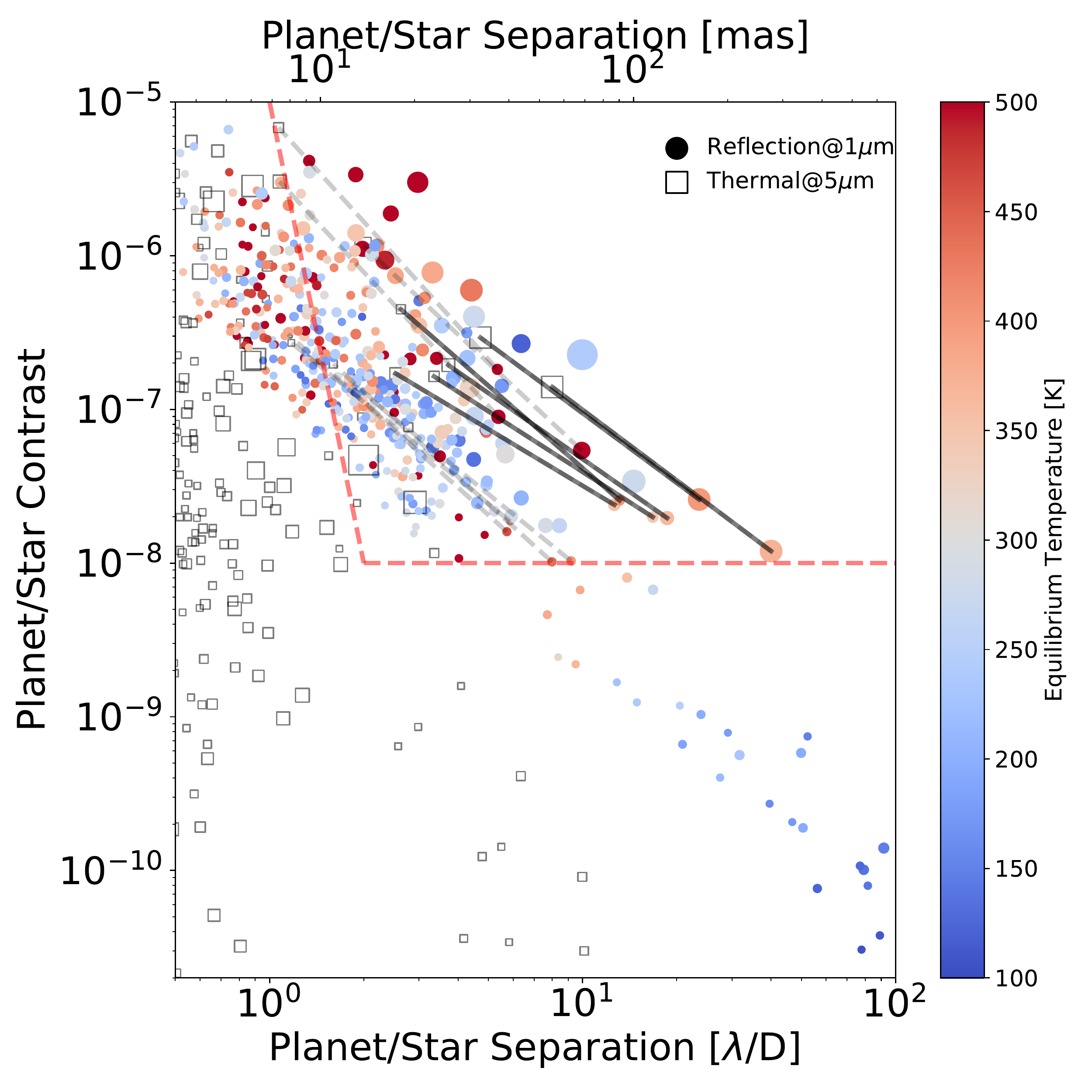}
    \caption{A simulated population of rocky and ice-giant planets (R$<$4 R$_{\rm{Earth}}$) within 10 pc based on Kepler statistics~\citep{dressing&charbonneau15, Petigura2013}. Filled circles are small planets as seen in reflected light (albedo=0.3) at 1 $\mu$m by a 30-m telescope with color indicating equilibrium temperature and size indicating distance from the solar system. Most amenable systems for detection in reflected light are planets around M stars. Open squares are small planets as seen at 5 $\mu$m. The most amenable systems for detection in thermal emission are small planets around hotter (FGK) stars. Red dashed line indicates nominal parameter space that can be potentially probed by direct imaging at ELTs. Linked pairs are planetary systems that can be potentially detected in both reflected light and thermal emission with line grey scale representing detection difficulty (the darker, the more amenable in terms of angular distance and contrast). Note that all planets shown in linked pairs have R$>$2 R$_{\rm{Earth}}$ and are around solar-type stars.}
    \label{fig:landscape}
\end{wrapfigure}

The radial velocity (RV) and photometric transit techniques have led to discoveries of thousands of planets outside the solar system (exoplanets), painting a detailed picture of exoplanet demographics within $\sim$1 AU of host stars.  Our understanding of planets at longer periods, which may well dominate in total numbers (but where RV and transit are much less effective) is very limited.   In this regime, direct imaging serves as a powerful tool for discovery and characterization.  Direct imaging and spectroscopy permit (1) detection of small planets including Earth-sized planets, super-earths, and Neptune-sized planets; (2) determination of luminosity, equilibrium temperature, radius, and atmospheric composition; (3) full accounting of the energy budget and climate for planets detected in both reflected light and thermal emission; and (4) planetary rotation and atmospheric dynamics when high resolution spectroscopy is feasible.   TMT and GMT will extend this approach to unprecedented ranges of small planet size and clement equilibrium temperatures, including the liquid-water ``habitable zone", enabling us, for the first time, to place the Solar System and Earth in context. {\bf{A survey of the nearest (within several pc) small ($<$4 R$_{\oplus}$) exoplanets in both thermal emission and reflected light has the potential to transform our knowledge of these worlds (Fig. \ref{fig:landscape}).}}
\newline
\newline
Both the reflected light and thermal emission of a planet depend on stellar luminosity, orbital separation, planet radius, and albedo as a function of wavelength:  the amount of reflected light is high when the albedo is high while the thermal emission is highest when the albedo is low.  Planets in thermal equilibrium with their host star emit steady amounts of infrared radiation regardless of age.  Wavelengths $<$ 2.3 $\mu$m are best suited for imaging mature planets in reflected light while thermal emission usually dominates from 3-14 $\mu$m. These approaches are complementary and, as we discuss below, when studying planets with both techniques, the whole is greater than the sum of the parts. 

\section{Thermal Imaging of Planets around Solar-type Stars}
\label{sec:thermal}

Observations at 3-14 $\mu$m are best-suited for characterizing the emitting temperature and intrinsic luminosity of planets with temperatures of 270-800 K, including those in a ``habitable zone" where liquid water can exist on solid surfaces.  Although observations from space (with \emph{JWST}) benefit from a much lower thermal background, 30-m ELTs equipped with adaptive optics and diffraction suppression optics should out-perform the 6-m \emph{JWST} in terms of contrast limit within $10\lambda/D$~\citep{Meyer2018}, enabling detection of dozens of small planets.  Most thermal emission detection will occur in the L- and M-bands, including terrestrial, super-earth, and Neptune-sized objects inside the habitable zone \citep{Quanz2015}.  Such observations are enabled by high contrast imagers that span 0.8-14 $\mu$m and high-resolution AO-fed spectrographs on the ELTs. 

\begin{figure}[h!]
    \centering
    \includegraphics[width=0.7\textwidth]{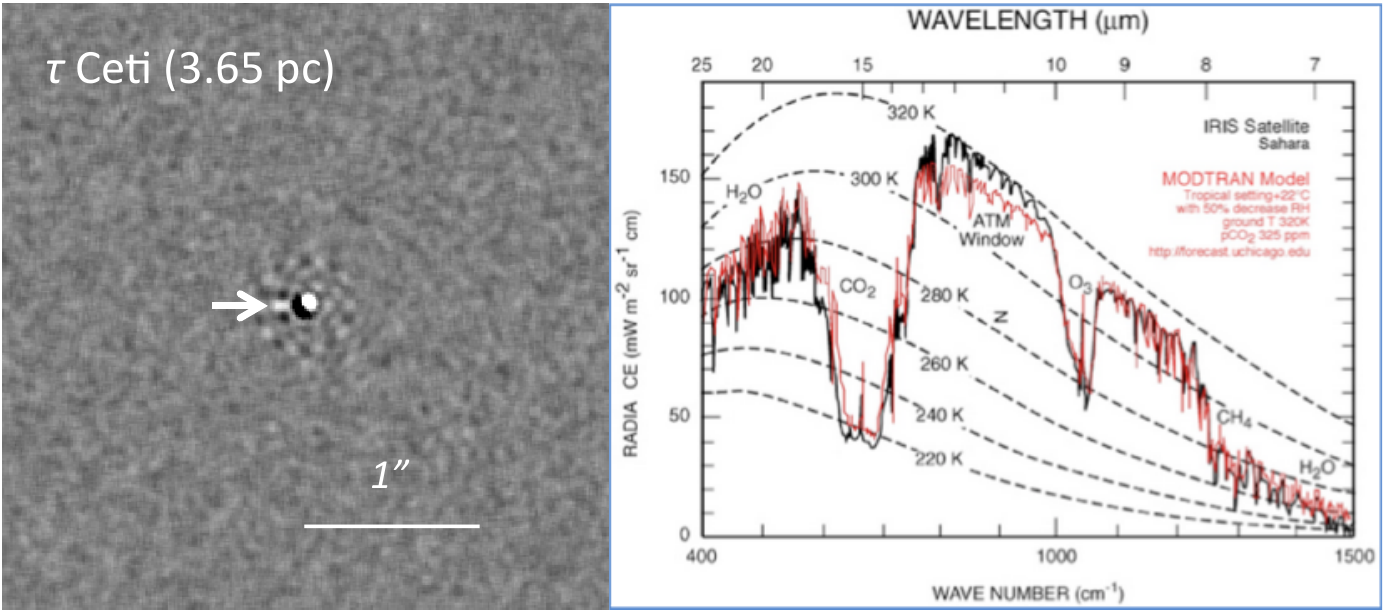}
    \caption{Thermal infrared (10 $\mu$m) imaging and characterization.  Left:  Simulated image of RV candidate $\tau$ Ceti e (1.3 R$_{\rm{Earth}}$ planet at 0.5 AU, as marked) in equilibrium with the host star at the inner edge of the "habitable zone".  Scale bar is 1$^{\prime\prime}$ for reference.  A coronagraph and advanced post-processing techniques are used to suppress the central star. Right: Observed and modeled spectrum of Earth at 7-13 $\mu$m showing major absorption features due to atmospheric constituents.
    \label{fig:thermal}
    }
\end{figure}

Detection and characterization of Earth analogs in the habitable zone of the nearest Sun-like stars will be enabled by the angular resolution and thermal IR ($\sim$10 $\mu$m) sensitivity of ELTs equipped with AO systems. The habitable zone will be resolved for targets within $d \approx$5 pc though, in the background limit, the required integration time scales as $d^4$.  The simulation in Fig. \ref{fig:thermal} for a target at 3.65 pc would take about 30 hours to reach a signal-to-noise $\sim$ 20 at a resolving power of $\sim$ 5.  Key targets in both the northern and southern hemispheres include the $\alpha$ Cen triple system, Procyon A, Altair, Sirius A, $\epsilon$ Eri, and $\tau$ Ceti, among others within 4 pc.  This implies the most distant accessible targets would take 10s of hours to detect at 10 $\mu$m.  There is also  significant synergy between these observations and a long-term extreme precision RV survey of possible targets which could refine our search~\citep{Kane2018}.
\newline
\newline
The abundance of key molecular species (CH$_4$, H$_2$O, CO$_2$, CO, HCN, and NH$_3$, see Fig. \ref{fig:thermal_reflection_detection}) in the atmospheres of planets 1 $< R_{\rm{Earth}} < $ 4 can be retrieved using spectra with resolution $\lambda/\Delta \lambda \sim$100 from 3-14 $\mu$m.   These abundances are crucial to understanding the origin and evolution of planetary atmospheres (both primary due to accretion and secondary due to delivery as well as out-gassing) including the potential formation of pre-biotic compounds. The 3-5 $\mu$m window is vital for accurate constraints on the C/O ratio~\citep[e.g., ][]{Oberg2011, Chapman2017} while the 10 $\mu$m window provides information on NH$_3$ and O$_3$ (a proxy for molecular O$_2$).  Broad wavelength coverage will permit detailed studies of cloud properties (particle size and composition) as the optical properties are expected to transition from gray to wavelength dependent in the thermal-IR.   % E.G.: size-dependent, of course: reference here?     
Neither \emph{JWST} nor \emph{WFIRST-CGI} will be able to directly image an Earth-like planet around a Sun-like star.  %Planned NASA missions (e.g., JWST and WFIRST CGI) will not enable this landmark discovery.  
Further, thermal infrared observations of rocky exoplanets could significantly enhance the science return of potential future NASA direct imaging missions (e.g., \emph{LUVOIR} or \emph{HABEX}), e.g., detecting small planets prior to these NASA missions, and providing complementary wavelengths for planet characterization.  % at a fraction of the cost. 

\section{Detecting Reflected-Light from Planets Around M Stars}\label{sec:reflection}

Current exoplanet detection techniques overwhelmingly favor detection of planets around M stars: RV and transit signals are larger because the stars are less massive and smaller, and planets with a given equilibrium temperature will have shorter orbital periods, with concomitantly larger  transit probabilities and RV signals.  There appear to be more small planets, at least on close-in orbits, around M stars \citep[2.5 planets per star, ][]{dressing&charbonneau15}. Many systems such as Proxima Cen b~\citep{Escude2016} and Barnard's star b~\citep{Ribas2018} will be detected in the future with the advent of extreme precision RV spectrographs in both optical and NIR wavelengths.  Direct imaging followup to RV-selected stars is efficient as the existence of a planet is already very likely and the search space is constrained by RVs~\citep{Kane2018}.     
\newline
\newline
These factors motivate a two-part strategy for detecting M dwarf planets: (1) direct imaging \emph{follow-up} to confirm planet candidates previously identified by the RV method, and (2) a ``blind" survey. For the first part, the combination of RV and direct imaging will resolve the geometric ($\sin i$) ambiguity in RV determination of planetary mass.  Combined with a radius estimate from direct imaging (modulo an assumption of albedo) the planet's mean density can be determined and overall composition (i.e. rocky like Earth or gas- and ice-rich like Neptune) inferred.  The range of planet radii probed by this survey spans Earth to  Neptune where planets can have a range of bulk compositions \citep{WeissMarcy2014}.   
\newline
\newline
Nevertheless, some small planets are difficult to detect by RV because (1) they are further away and RV signal is weaker; and (2) $\sim$50\% of M dwarfs are fast-rotating and therefore not amenable to the RV method~\citep{Newton2016}. In this case, a ``blind" survey of those M dwarfs that are most amenable to detection of direct detection of small planets.  This expands the planet search to $\sim$20 additional single M stars within 5~pc.  Spectroscopy of detected planets will unambiguously reveal the diversity of atmospheres for  these planets, with size ranging from Earth to Neptune and with mild temperature ranging from 200-500 K.  
\newline
\newline
The discovery space opened by reflected-light direct imaging will provide legacy value by enabling several high value investigations, including: (1) Measuring  photometric and spectroscopic variations with planet rotation to determine rotation rate and reveal surface features such as cloud coverage and/or land/ocean distribution~\citep{Fujii2012, crossfield14};  (2) High-dispersion spectroscopy, when combined with sophisticated atmospheric models, will provide constraints on a planet's atmospheric chemistry and reveal potential biosignatures~\citep{wang_etal17}. (3) At high spectral resolution, we can measure the radial velocity of an exoplanet and therefore break the mass-inclination degeneracy of host star RV studies \citep{Piskorz2017};  (4) Spectral line profiles will provide insights into rotation rates not discernable from (1) above ~\citep{Snellen2014,Bryan2018} as well as  atmosphere dynamics, e.g., circulation and vertical structure~\citep{Kempton2014}.  

% If a sufficiently large sample is obtained, comparison of the demographics of planets around active, rapidly rotating vs. quiet, slowly-rotating stars may constrain the impact of stellar environment on planet climate (and perhaps habitability).  We can also compare the results of our survey to those from the sample of the comparable planets around higher mass AFGK stars to explore emerging trends.   

\section{Towards Comprehensive Understanding of Other Worlds}
\label{sec:holy_grail_pl}

\begin{figure}[ht]
 \centering
 \includegraphics[width=0.95\linewidth]{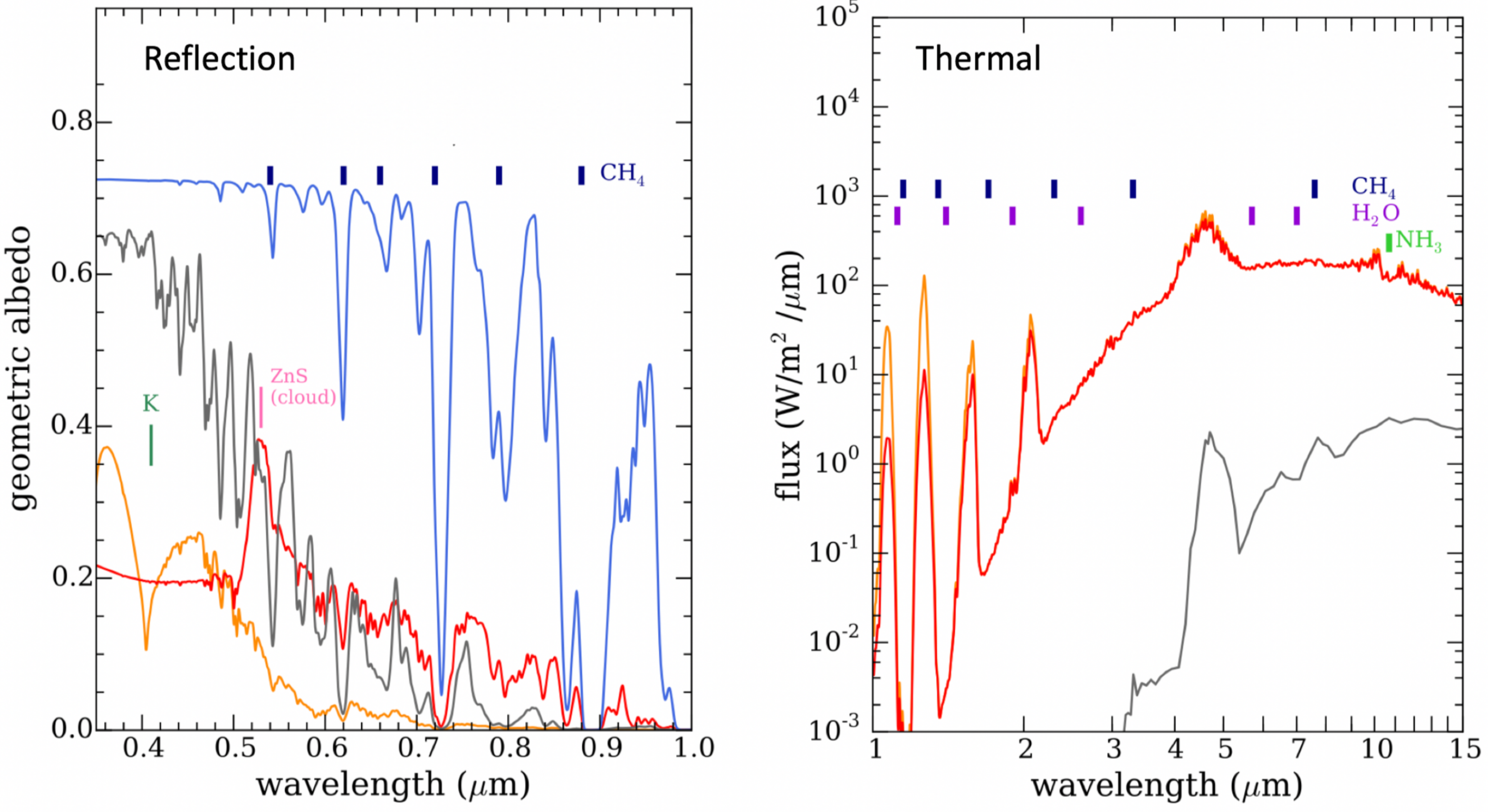}
 \qquad
 \caption{Left: diversity of reflection spectra for a Neptune-sized planets with a range of mild temperatures (200-500 K). Right: same as left but in thermal emission. The transition wavelength from reflected light dominance to thermal emission dominance is at $\sim$0.9 $\mu$m. Direct imaging and spectroscopy can probe the diversity of chemical compositions and cloud properties. More importantly, detection and characterization of small planets in both reflection and thermal wavelengths allows for a study of an ``energy budget'' -- the basis of planetary climate. 
 \label{fig:thermal_reflection_detection}}
\end{figure}

For the nearest, brightest super-Earths and ice giants, we may be able to perform deep observations at multiple epochs at both visible and thermal infrared wavelengths (linked pairs in Fig. \ref{fig:landscape}). This permits better characterization of their physical and atmospheric properties, and specifically allow an ``energy budget'' -- the basis of planetary climate -- to be constructed.  First, observations of the planet at multiple positions along the orbit (along with any RV measurements) we will better constrain the orbit and phase angle at each epoch.  Second, by measuring the spectral intensity of planets over wavelength ranges that span the peaks in emission we will determine their  integrated fluxes, effective temperatures (T$_{eff}$), and radii \citep{Traub1}.  Observations at visible wavelengths, combined with knowledge of the planet radius and phase angle will then constrain the overall scattering properties of the atmosphere, including clouds \citep{Nayak}. While the scattering phase function of cloud particles gives us information on the size of the particles and their index of refraction, additional information on their composition can be drawn from multi-wavelength observations covering a wide spectral range.  Combining the radius of the planet and stellar irradiance with the integrated reflected light yields the Bond albedo, which can be used to calculate an equilibrium temperature (T$_{\rm eq}$), for comparison to the observed $T_{\rm eff}$ for consistency, or indication of an internal energy source  such as the primordial heat of accretion \citep{Traub1}.  Taken together, all these measurements will lead to a deeper and more comprehensive understanding of Earth- to Neptune-size exoplanets (Fig. \ref{fig:thermal_reflection_detection}): their size distribution, composition, atmospheric compositions, and climates, arguably all necessary in order to understand diversity and potentially habitability.  Such detailed studies will only be possible with a comprehensive US ELT Program, enabling us, perhaps for the first time, to detect an active greenhouse and complex organic chemistry, on a planet outside our solar system, and begin to place our world in context.

\section{The Enabling Power of a Robust US ELT Program}

The observations described here are enabled by ELTs and are not possible with existing or planned ground- or even space-based telescopes.  The required angular resolution, and sensitivity, provided by ELTs is enabling for both thermal emission and reflected light studies of small planets (1 $< R <$ 4 R$_{\rm{Earth}}$) interior to, within, and beyond the habitable zone.  These planets may dominate global exoplanet demographics and contain the clues we need to understand planet formation, as well as complex chemical evolution, including that needed to promote the biochemical origins of life.  Given the modest numbers of touchstone targets, the possibilities for complementary observations from both hemispheres and all-sky access, a robust US ELT Program is required for us to take these next steps, required to place our solar system in context. 

%----------------------------------------------------------------------------------------
%	REFERENCE LIST
%----------------------------------------------------------------------------------------

%\bibliographystyle{unsrt}

\begin{multicols}{2}

\bibliographystyle{apj}
\setlength{\bibsep}{0.0pt}
\scriptsize
\bibliography{literature}

\newcommand{\noopsort}[1]{}
\begin{thebibliography}{20}
\expandafter\ifx\csname natexlab\endcsname\relax\def\natexlab#1{#1}\fi

\bibitem[{{Anglada-Escud{\'e}} {et~al.}(2016){Anglada-Escud{\'e}}, {Amado},
  {Barnes}, {Berdi{\~n}as}, {Butler}, {Coleman}, {de La Cueva}, {Dreizler},
  {Endl}, {Giesers}, {Jeffers}, {Jenkins}, {Jones}, {Kiraga}, {K{\"u}rster},
  {L{\'o}pez-Gonz{\'a}lez}, {Marvin}, {Morales}, {Morin}, {Nelson}, {Ortiz},
  {Ofir}, {Paardekooper}, {Reiners}, {Rodr{\'{\i}}guez},
  {Rodr{\'{\i}}guez-L{\'o}pez}, {Sarmiento}, {Strachan}, {Tsapras}, {Tuomi}, \&
  {Zechmeister}}]{Escude2016}
{Anglada-Escud{\'e}}, G., {et~al.} 2016, \nat, 536, 437

\bibitem[{{Bryan} {et~al.}(2018){Bryan}, {Benneke}, {Knutson}, {Batygin}, \&
  {Bowler}}]{Bryan2018}
{Bryan}, M.~L., {Benneke}, B., {Knutson}, H.~A., {Batygin}, K., \& {Bowler},
  B.~P. 2018, Nature Astronomy, 2, 138

\bibitem[{{Chapman} {et~al.}(2017){Chapman}, {Zellem}, {Line}, {Vasisht},
  {Bryden}, {Willacy}, {Iyer}, {Bean}, {Cowan}, {Fortney}, {Griffith},
  {Kataria}, {Kempton}, {Kreidberg}, {Moses}, {Stevenson}, \&
  {Swain}}]{Chapman2017}
{Chapman}, J.~W., {et~al.} 2017, PASP, 129, 104402

\bibitem[{{Crossfield}(2014)}]{crossfield14}
{Crossfield}, I.~J.~M. 2014, \aap, 566, A130

\bibitem[{{Dressing} \& {Charbonneau}(2015)}]{dressing&charbonneau15}
{Dressing}, C.~D., \& {Charbonneau}, D. 2015, \apj, 807, 45

\bibitem[{{Fujii} \& {Kawahara}(2012)}]{Fujii2012}
{Fujii}, Y., \& {Kawahara}, H. 2012, \apj, 755, 101

\bibitem[{{Kane} {et~al.}(2018){Kane}, {Meshkat}, \& {Turnbull}}]{Kane2018}
{Kane}, S.~R., {Meshkat}, T., \& {Turnbull}, M.~C. 2018, \aj, 156, 267

\bibitem[{{Kempton} {et~al.}(2014){Kempton}, {Perna}, \& {Heng}}]{Kempton2014}
{Kempton}, E.~M.-R., {Perna}, R., \& {Heng}, K. 2014, \apj, 795, 24

\bibitem[{{Meyer} {et~al.}(2018){Meyer}, {Currie}, {Guyon}, {Hasegawa},
  {Kasper}, {Marois}, {Monnier}, {Morzinski}, {Packham}, \&
  {Quanz}}]{Meyer2018}
{Meyer}, M.~R., {et~al.} 2018, arXiv e-prints: 1804.03218

\bibitem[{{Nayak} {et~al.}(2017){Nayak}, {Lupu}, {Marley}, {Fortney},
  {Robinson}, \& {Lewis}}]{Nayak}
{Nayak}, M., {Lupu}, R., {Marley}, M.~S., {Fortney}, J.~J., {Robinson}, T., \&
  {Lewis}, N. 2017, \pasp, 129, 034401

\bibitem[{{Newton} {et~al.}(2016){Newton}, {Irwin}, {Charbonneau},
  {Berta-Thompson}, {Dittmann}, \& {West}}]{Newton2016}
{Newton}, E.~R., {Irwin}, J., {Charbonneau}, D., {Berta-Thompson}, Z.~K.,
  {Dittmann}, J.~A., \& {West}, A.~A. 2016, \apj, 821, 93

\bibitem[{{{\"O}berg} {et~al.}(2011){{\"O}berg}, {Murray-Clay}, \&
  {Bergin}}]{Oberg2011}
{{\"O}berg}, K.~I., {Murray-Clay}, R., \& {Bergin}, E.~A. 2011, ApJL, 743, L16

\bibitem[{{Petigura} {et~al.}(2013){Petigura}, {Howard}, \&
  {Marcy}}]{Petigura2013}
{Petigura}, E.~A., {Howard}, A.~W., \& {Marcy}, G.~W. 2013, Proceedings of the
  National Academy of Science, 110, 19273

\bibitem[{{Piskorz} {et~al.}(2017){Piskorz}, {Benneke}, {Crockett}, {Lockwood},
  {Blake}, {Barman}, {Bender}, {Carr}, \& {Johnson}}]{Piskorz2017}
{Piskorz}, D., {et~al.} 2017, \aj, 154, 78

\bibitem[{{Quanz} {et~al.}(2015){Quanz}, {Crossfield}, {Meyer}, {Schmalzl}, \&
  {Held}}]{Quanz2015}
{Quanz}, S.~P., {Crossfield}, I., {Meyer}, M.~R., {Schmalzl}, E., \& {Held}, J.
  2015, International Journal of Astrobiology, 14, 279

\bibitem[{{Ribas} {et~al.}(2018){Ribas}, {Tuomi}, {Reiners}, {Butler},
  {Morales}, {Perger}, {Dreizler}, {Rodr{\'{\i}}guez-L{\'o}pez}, {Gonz{\'a}lez
  Hern{\'a}ndez}, {Rosich}, {Feng}, {Trifonov}, {Vogt}, {Caballero}, {Hatzes},
  {Herrero}, {Jeffers}, {Lafarga}, {Murgas}, {Nelson}, {Rodr{\'{\i}}guez},
  {Strachan}, {Tal-Or}, {Teske}, {Toledo-Padr{\'o}n}, {Zechmeister},
  {Quirrenbach}, {Amado}, {Azzaro}, {B{\'e}jar}, {Barnes}, {Berdi{\~n}as},
  {Burt}, {Coleman}, {Cort{\'e}s-Contreras}, {Crane}, {Engle}, {Guinan},
  {Haswell}, {Henning}, {Holden}, {Jenkins}, {Jones}, {Kaminski}, {Kiraga},
  {K{\"u}rster}, {Lee}, {L{\'o}pez-Gonz{\'a}lez}, {Montes}, {Morin}, {Ofir},
  {Pall{\'e}}, {Rebolo}, {Reffert}, {Schweitzer}, {Seifert}, {Shectman},
  {Staab}, {Street}, {Su{\'a}rez Mascare{\~n}o}, {Tsapras}, {Wang}, \&
  {Anglada-Escud{\'e}}}]{Ribas2018}
{Ribas}, I., {et~al.} 2018, \nat, 563, 365

\bibitem[{{Snellen} {et~al.}(2014){Snellen}, {Brandl}, {de Kok}, {Brogi},
  {Birkby}, \& {Schwarz}}]{Snellen2014}
{Snellen}, I.~A.~G., {Brandl}, B.~R., {de Kok}, R.~J., {Brogi}, M., {Birkby},
  J., \& {Schwarz}, H. 2014, \nat, 509, 63

\bibitem[{{Traub} \& {Oppenheimer}(2010)}]{Traub1}
{Traub}, W.~A., \& {Oppenheimer}, B.~R. 2010, {Direct Imaging of Exoplanets},
  ed. S.~{Seager}, 111--156

\bibitem[{{Wang} {et~al.}(2017){Wang}, {Mawet}, {Ruane}, {Hu}, \&
  {Benneke}}]{wang_etal17}
{Wang}, J., {Mawet}, D., {Ruane}, G., {Hu}, R., \& {Benneke}, B. 2017, AJ, 153,
  183

\bibitem[{{Weiss} \& {Marcy}(2014)}]{WeissMarcy2014}
{Weiss}, L.~M., \& {Marcy}, G.~W. 2014, \apjl, 783, L6

\end{thebibliography}
\end{multicols}

\pagebreak

\end{document}